*Juggling too many balls at once: Qualitatively different effects when measuring priming and masking in single, dual, and triple tasks*

*arXiv:2208.13675 [q-bio.NC]*

Melanie Biafora & Thomas Schmidt
University of Kaiserslautern, Germany

Corresponding author:
Dr. Melanie Biafora M.Ed.
University of Kaiserslautern
Faculty of Social Sciences
Experimental Psychology Unit
Erwin-Schrödinger Straße Geb. 57
67663 Kaiserslautern
Germany

Tel.: +49 (0)631 250 3940
E-Mail: melanie.biafora@sowi.uni-kl.de

**Abstract**

Dissociation paradigms examine dissociations between *indirect measures* of prime processing and *direct measures* of prime awareness. It is debated whether direct measures should be *objective* or *subjective*, and whether these measures should be obtained on the same or separate trials. In two metacontrast experiments, we measured prime discrimination, PAS ratings, and response priming either separately or in multiple tasks. Single tasks show the fastest responses in priming and therefore most likely meet the assumption of feedforward processing as assumed under Rapid-Chase Theory. Similarly, dual tasks allow for a fast response activation by the prime; nevertheless, prolonged responses and slower errors occur more often. In contrast, triple tasks have a negative effect on response activation: responses are massively slowed and fast prime-locked errors are lost. Moreover, decreasing priming effects and prime identification performance result in a loss of a double dissociation. Here, a necessary condition for unconscious response priming, *feedforward processing*, is violated.

**Introduction**

In investigating perception without awareness, response priming is a powerful tool to dissociate visual awareness of a stimulus from its visuomotor effects (Klotz & Neumann, 1999; Klotz & Wolff, 1995; Vorberg, Mattler, Heinecke, Schmidt, & Schwarzbach, 2003). In response priming, participants perform speeded responses to identify a target stimulus (e.g., as a diamond or square) which is preceded by a briefly presented prime (e.g., another diamond or square) that is either mapped to the same response as the target (*consistent trials*) or to the alternative response (*inconsistent trials*). At stimulus-onset asynchronies (SOAs) between prime and target of up to about 100 ms, consistent primes speed responses to the target, whereas inconsistent primes slow responses and provoke errors. These priming effects in response times and error rates increase with prime-target SOA (Vorberg et al., 2003). Several lines of evidence indicate that this type of response conflict occurs because the first visuomotor activity is controlled by the prime alone. (1) Error rate increases with SOA in inconsistent trials while few errors occur in consistent trials; (2) errors are as fast as the fastest correct responses; and (3) the fastest responses always follow the identity of the prime (i.e., they are always correct when the prime is consistent and always incorrect when it is inconsistent; Panis & Schmidt, 2016). Moreover, (4) for inconsistent trials the early time-course of priming effects in pointing responses, force profiles, and lateralized readiness potentials is invariant, strongly suggesting that these processes are initially controlled exclusively by the prime (Eimer & Schlaghecken, 1998; Klotz, Heumann, Ansorge, & Neumann, 2007; F. Schmidt, Weber, Schmidt, 2014; T. Schmidt, Niehaus, & Nagel, 2006; T. Schmidt & Schmidt, 2009; Vath & Schmidt, 2007; Verleger, Jaśkowski, Aydemir, van der Lubbe, & Groen, 2004).

These observations lead to the proposal that in response priming, the prime activates the response in a simple feedforward process before the target is able to take over, and that the longer the SOA, the longer the prime is able to control the response on its own (*Rapid-Chase Theory*, T. Schmidt et al., 2006, 2011; see also T. Schmidt, 2014).

The time-course of response priming can be qualitatively dissociated from the time-course of visual awareness for the prime. A special type of backward masking called *metacontrast* can be used to generate prime discrimination functions that either increase with SOA (*type-A masking*) or decrease with SOA (*type-B masking*; Breitmeyer & Öğmen, 2006). Under conditions of metacontrast masking, the target itself serves as a metacontrast mask for the prime and for prime-mask SOAs up to about 100 ms, response priming can increase with SOA no matter whether the ability to discriminate the prime is virtually perfect, virtually absent, increases, or decreases with SOA (Mattler, 2003; F. Schmidt, Haberkamp, & Schmidt, 2011; T. Schmidt & Vorberg, 2006; Vorberg et al., 2003).[1] Rapid-Chase Theory assumes that such dissociations between priming and prime discriminability are made possible by the feedforward nature of response priming: Whereas priming effects are based on a simple feedforward sweep of visuomotor processing, visual awareness of the prime depends on recurrent processing loops forming while the response is already being activated (Bullier, 2001; Lamme & Roelfsema, 2000; F. Schmidt et al., 2011; VanRullen & Koch, 2003). Therefore, response activation by the prime remains intact even though the

(backward) mask interrupts the feedback signal necessary for visual awareness (Di Lollo, Enns, & Rensink, 2000; Fahrenfort, Scholte, & Lamme, 2007; Lamme, Zipser, & Spekreijse, 2002).

This comparison is an example of the *dissociation paradigm*, which seeks to establish dissociations between measures of visual awareness of a critical stimulus (*direct measures*) and measures indicating that the critical stimulus has been processed at all (*indirect measure*, Reingold & Merikle, 1988; also see Erdelyi, 1986; Reingold, 2004; Reingold & Ray, 2006). Usually, this procedure is employed to establish a *simple dissociation* between direct and indirect measures – e.g., significant priming effects in the indirect measure while a direct measure of awareness for the prime is at chance level. Unfortunately, this traditional criterion for perception without awareness requires the strong assumption that the measure of awareness is *exhaustive* with respect to any relevant conscious information. Without an exhaustiveness assumption, zero sensitivity in the direct measure is not conclusive because the measure may have just failed to pick up some of the conscious information (Eriksen, 1960; Holender, 1986; Reingold & Merikle, 1988; T. Schmidt, 2008; T. Schmidt & Vorberg, 2006). A much more informative data pattern is a *double dissociation*, which is the demonstration that some experimental manipulation leads to opposite changes in direct and indirect measures – e.g., that an increase in prime-mask SOA leads to an increase in response priming but a *decrease* in prime discrimination performance, as observed by Vorberg et al. (2003). Double dissociations require only mild measurement assumptions and provide direct evidence that the ability to discriminate the prime cannot explain priming (T. Schmidt & Vorberg, 2006). We have recently proposed an experimental technique for facilitating double dissociations (*induced dissociations*; Biafora & Schmidt, 2019).

There is an ongoing debate about how awareness should be measured in the dissociation paradigm. This debate mostly distinguishes between two types of direct measures: *objective measures*, for which participants' responses can be compared with the actual stimulus and are therefore classifiable as correct or incorrect (e.g., yes-no detection or discrimination; two-alternative forced choice), and *subjective measures* where participants report an internal state that cannot be validated externally (e.g., ratings on stimulus brightness, clarity of impression, or confidence in correct identification; Cheesman & Merikle, 1984, 1986; Reingold, 2004). Even though some authors hold strong views that only one or the other class of measures should be considered a "gold standard", it is clear that different types of direct tasks not only differ in their object of measurement (e.g., the ability to distinguish shapes, the subjective confidence in the decision) but also in their *criterion content* – i.e., the information that the observer is actually using when performing the task (Kahneman, 1968).

In psychophysical procedures, objective and subjective measures are often used jointly, e.g., when constructing a receiver-operating characteristic that plots objective hit and false alarm rates as a function of subjective confidence ratings (ROC-curves). Several authors argue that subjective and objective measures can be equally sensitive, because they found that when participants report that subjective visibility is absent, their performance in an objective discrimination task is also at chance (e.g., Avneon & Lamy, 2018; Lamy, Alon, Carmel, & Shalev, 2015, Lamy, Carmel, & Peremen, 2017; Peremen & Lamy, 2014; Ramsøy & Overgaard, 2004).

Another issue is whether direct and indirect measures should be obtained in separate trials (blocks, sessions) or on the same trial. Peremen and Lamy (2014) argue that the double dissociations observed between response priming and prime discrimination performance (increasing priming effects under decreasing prime discrimination; Vorberg et al., 2003) are the result of a measurement artifact: because priming and prime discrimination are measured in separate blocks, observers attend to the prime in one task and to the target in the other (Van den Bussche et al., 2013). Peremen and Lamy (2014) asked participants to give three responses in each trial: first a speeded response to the mask, next an unspeeded discrimination response to the prime, and finally a visibility rating on a four-point *Perceptual Awareness Scale* (*PAS*, Ramsøy & Overgaard, 2004). They observed a u-shaped masking function in both discrimination performance and PAS ratings, but also a u-shaped time-course of the priming effect. This pattern was replicated in two control experiments that employed dual instead of triple tasks: one involving prime discrimination and PAS ratings, and one involving prime discrimination and speeded responses to the mask. In a final experiment, participants responded only to the mask, and now priming effects simply increased with

SOA, just as in Vorberg et al.'s (2003) original study. The authors conclude that the double dissociation pattern with increasing priming effects under decreasing prime discrimination is an artifact of using single tasks instead of dual or triple tasks.

While in single tasks priming and prime discrimination are performed in separate blocks or sessions, observers can focus on the task relevant stimulus (either the mask which simultaneously serves as target or the prime, respectively). In dual- or triple tasks, in contrast, observers are forced to hold all stimuli in memory because they need to perform more than one task within the same trial (e.g., priming and prime discrimination in the dual task and some additional rating tasks under triple-task conditions) on the basis of the same memory representation. For this reason, average response times under multiple tasks can be expected be longer than under conditions of single response priming tasks (In the study of Lamy, Peremen and Carmel (2017) a table of absolute average response times reveals that responses are about 150 ms longer under dual tasks than usually obtained under single tasks.

Thus, can dual or triple tasks still provide the necessary condition for unconscious response priming, namely rapid feedforward processing without conscious control, even if each task type operates with a different cognitive load? If responses are no longer activated automatically by simple feedforward processing, priming effects may no longer increase over SOA. Rather response inhibition can cause decreasing priming effects (Eimer & Schlaghecken, 1998; Panis & Schmidt, 2016; T. Schmidt, Hauch, & Schmidt, 2015). Any dissociation between an indirect measure of prime processing and a direct measure of prime discrimination can be lost, and the meaning of the dissociation paradigm becomes invalid.

Furthermore, single and multiple tasks should differ in the amount of *sensory attention* directed to the task-relevant stimuli. If the prime is the only task-relevant stimulus while the mask can be ignored, the task is optimized for directing all attentional resources to the prime. This should facilitate prime identification as compared to a single mask-identification task where the prime can be ignored. (Note that prime identification performance is traditionally measured under optimal conditions out of a desire for a conservative test of unconscious processing). In a multitask setting, sensory attention has to be split between the prime and mask, rendering any finding of low prime identification performance less convincing.

There are several plausible sources for a delay in response times and an increase of mental load under dual or triple tasks. In a multiple-task setting, divided attention between the tasks places higher demands on *cognitive control* (Pashler, 1989, 1994b) that could result in slower reaction times or impaired performances in prime identification (e.g., switch costs, task-shift costs, and task-set reconfiguration costs; Kiesel, Wendt, & Peters 2007; Pashler, Johnston, & Ruthruff, 2001; Waszak, Hommel, & Allport, 2003). In tasks where two stimuli are presented in rapid succession and participants have to give speeded responses to each of them, the second response is systematically slowed as the SOA between the stimuli decreases, probably because of interference in response selection (*psychological refractory period*; Pashler, 1989, 1994a, 1994b). This raises the question whether response selection in the second task (prime discrimination) interferes with response selection in the first task (mask discrimination in the priming task) because they involve the same two responses. If so, the second task could directly interfere with response priming.

Furthermore, multiple tasks lead to a high *working memory load* because they require observers to simultaneously store the identity of the mask the identity of the prime, and the information required by the PAS. Working memory load, in turn, interacts with visual attention (Olivers, Meijer, & Theeuwes, 2006, Olivers, Peters, Houtkamp, & Roelfsema, 2011). This raises the question whether those memory contents remain independent of each other, or whether responses in later tasks are influenced by responses given earlier (e.g., once identification responses have been made to the masks and primes, subjective awareness might seem stronger in hindsight than it originally was).

In this paper, we want to address the following sets of questions, which are still part of debate: *First*, are objective and subjective measures of awareness equally valid for measuring awareness of the critical prime feature? *Second*, is it important for the logic of dissociation paradigm whether direct and indirect measures are applied together within the same trial (multiple task) or separately (single task)? Generally and most importantly, can multiple tasks guarantee fast forward processing to the same extend as single tasks, and thus provide a necessary condition for unconscious processing? To address these questions, we set up two experiments on response priming by simple geometric shapes (square vs. diamond) under metacontrast masking. In Experiment 1, we measured response priming, prime discrimination performance, and PAS ratings either under single-task conditions (with each task performed in a separate session) or under triple-task conditions (with each task performed within the same trial for three sessions). In Experiment 2, we contrasted single- and dual-task conditions, leaving out the PAS rating.

**Experiment 1**

This experiment was conducted to study the time course of response priming and prime visibility under metacontrast masking in single and triple tasks. We employed two measures of visibility: an objective measure (response accuracy in prime discrimination) and a subjective measure (visibility ratings on the Perceptual Awareness Scale, Ramsøy & Overgaard, 2004). On each trial of the triple task, participants first gave a speeded response to the shape of the mask (square or diamond), then tried to discriminate the prime (square or diamond) without time pressure, and finally gave a visibility rating on the PAS. The same tasks were also performed as single tasks in separate sessions. To obtain comparable conditions, we used the same stimuli, experimental processes, and participants for both task types.

If triple tasks require more cognitive control, higher memory load, and attention divided between stimuli, we expect some deterioration in performance relative to single tasks. In particular, responses to the mask should be slower, prime discrimination performance should be lower, and PAS ratings should be lower in triple than in single tasks. Furthermore, we suspected that priming effects under triple task conditions might no longer be carried by simple feedforward processing of the prime as maintained by Rapid-Chase Theory (T. Schmidt et al., 2006, 2011; see also T. Schmidt, 2014). We were therefore interested in whether the fast errors characteristic of a single priming task would still be observed under triple task conditions.

*Methods*
*Participants.* Eight right-handed volunteers, mainly students from the University of Kaiserslautern (4 male; mean age 29.1 years) took part in six 1-hour sessions. We tried to realize all sessions on separate days. However, in a few cases, two sessions were performed on the same day, with at least two hours between sessions. Each participant's vision was normal or corrected to normal. All of the participants were naïve to the purpose of the study. Participants were recruited in the course of bachelor theses, which is why attendees did not receive payment or course credit. Each of them gave informed consent and was treated according to the ethical guidelines of the American Psychological Association. After the final session, they were debriefed and received an explanation of the experiment.

*Apparatus.* The participants were seated in a dimly lit room in front of a color cathode-ray monitor (1280x1024 pixels, refresh rate 75 Hz) at a viewing distance of approximately 60 cm.

*Stimuli and Procedure.* We used prime and mask stimuli similar to those by Mattler (2003). At the beginning of each trial, a black fixation point appeared at the center of a white background (Fig. 1; ca. 63.0 cd/m). All stimuli were black squares or diamonds (0.04 cd/m²), differing in size and shape characteristics. Primes had an edge length of 1 cm (0.96° of visual angle) and appeared at fixation. Masks were squares or diamonds with an edge length of about 1.6 cm (1.53°) appearing at the same position as the primes. They had an additional central cut-out corresponding to the superposition of a square and a diamond prime, so that prime and mask shared adjacent but non-overlapping contours and both prime shapes could be masked by metacontrast (Breitmeyer & Öğmen, 2006; Di Lollo et al., 2000).

The experiment consisted of three different tasks that were either performed within the same trial (*triple-task* conditions), or separately in different sessions (*single-task* conditions). Participants performed a speeded mask identification task (*mID*), a non-speeded prime identification task (*pID*), and a visibility rating on a four-point PAS. With the mID task, we measured response priming in consistent and inconsistent trials as an *indirect* measure of prime processing. The pID task is an objective measure of prime discriminability, while the PAS is designed to be a subjective measure of general visibility of the prime. Both serve as *direct* measures of visual awareness of the prime.

Participants first performed one session of mID followed by one session of pID and finally one session of PAS ratings. Each session consisted of 31 blocks of 48 trials. Following this, participants performed three sessions in the triple-task condition, each consisting of 31 blocks of only 24 trials (to keep session durations similar between task conditions). We chose this sequence of tasks to give participants ample practice in the single tasks before taking on the more challenging triple task; and because we wanted a conservative estimate of the triple-task disadvantage we expected. The first block of each session was practice and not considered for further analysis. Stimulus sequences were identical in all sessions.

In the single-task conditions, each trial started with a central fixation point, followed by a prime presented for 27 ms that was either of the same shape as the mask (*consistent trial*) or the other shape (*inconsistent trial*). Finally, the mask appeared after a prime-mask SOA of 27, 40, 53, or 67 ms, and remained on screen until the response. The time interval from fixation onset to mask onset was constant at 600 ms. In the speeded *mask identification task* participants responded to the shape of the mask as quickly and correctly as possible by pressing button 'F' on the computer keyboard upon seeing a diamond or button 'J' upon seeing a square (or vice versa; the assignment was counterbalanced across participants). They used the two index fingers to respond and directly received visual feedback if the response was incorrect or too slow (> 1,000 ms). In the *prime identification task*, participants identified the shape of the prime without time pressure and without trial-to-trial feedback using the same stimulus-response mapping as in the mID task. After each block, participants could take a break and received summary feedback (on mean reaction time, mean accuracy, and number of errors in the mID, but only on mean accuracy in the pID). In the *perceptual awareness rating*, participants rated their visual impression of the prime on a four-point rating scale (PAS) by pressing the digit keys from 1 to 4. The specifications of the rating scale were visually presented on screen. The scale categories were "*1: kein Erlebnis*" *(no experience)*, "*2: flüchtiger Eindruck*" *(brief glimpse)*, "*3: nahezu klares Erlebnis*" *(almost clear experience)*, and "*4: klares Erlebnis*" *(clear experience)*". Additionally, participants had a brief description of the rating categories on the instruction sheet (Table 1; our German translation of the original PAS, Ramsøy & Overgaard, 2004). All combinations of prime shape, prime-mask consistency, and SOA were presented equiprobably and pseudo-randomly in each block.

Under *triple-task* conditions, participants performed exactly the same three tasks, but this time within the same trial. Therefore, participants first responded to the mask as quickly and accurately as possible; then tried to identify the prime (always using the same stimulus-response mapping); and finally rated their visual impression of the prime on the PAS. We changed timing according to the requirement that now three responses instead of one had to be given within a trial sequence. Therefore, the time interval from fixation to mask-onset was 1600 instead of 600 ms. The mask stimulus remained on screen until the final response. Visual feedback was given after the final response. To make the task easier, a visual instruction was presented at the time of prime identification, saying, "Identify the prime!". Participants received summary feedback after each block as described before.

*Table 1*: Categories and descriptions of the Perceptual Awareness Scale. *Left*: The original by Ramsøy and Overgaard, 2004; *right*: our translation into German.

| **Perceptual Awareness Scale** (Ramsøy & Overgaard, 2004) | | **Perceptual Awareness Scale** (our translation into German) | |
|---|---|---|---|
| Category | Description | Kategorie | Beschreibung |
| No experience | No impression of the stimulus. All answers are seen as mere guesses | 1 = "kein Erlebnis" (no experience) | Kein Eindruck des Stimulus. Meine abgegebene Antwort ist geraten. |
| Brief glimpse | A feeling that something has been shown. Not characterized by any content, and this cannot be specified any further | 2 = "flüchtiger Eindruck" (brief glimpse) | Ich hatte das Gefühl, dass irgendetwas gezeigt wurde, kann aber keine deutliche Unterscheidung treffen. |
| Almost clear experience | Ambiguous experience of the stimulus. Some stimulus aspects are experienced more vividly than others. A feeling of almost being certain about one´s answer | 3 = "nahezu klares Erlebnis" (almost clear experience) | Zweideutige Wahrnehmung des Stimulus. Einige Stimuluseigenschaften wurden klarer wahrgenommen als andere. Ich habe beinahe das Gefühl, den Stimulus erkannt zu haben. |
| Clear experience | Non-ambiguous experience of the stimulus. No doubt in one´s answer | 4 = "klares Erlebnis" (clear experience) | Kein zweideutiges Erlebnis des Stimulus, kein Zweifel bei der gegebenen Antwort. |

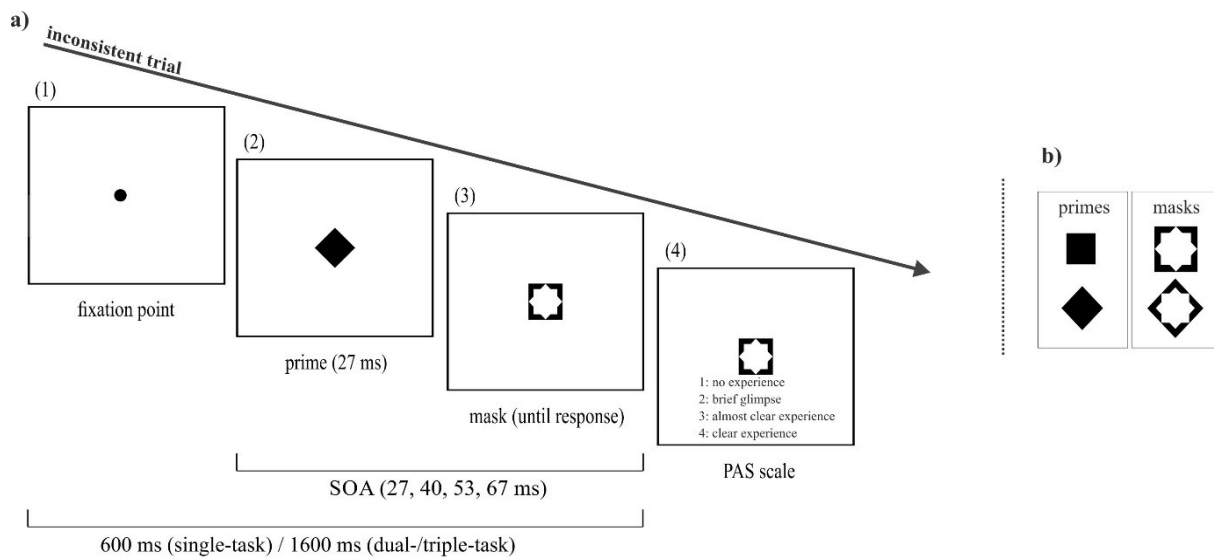

*Fig. 1.* (A) Stimuli and trial sequence in all experiments. The PAS display was only shown in the triple task and the respective single task of Experiment 1. (B) Prime and mask stimuli used in Experiments 1 and 2.

*Data treatment and statistical methods.* Dependent variables were response time and error rate in the mask identification task, response accuracy in the prime identification task, and the PAS ratings. Practice blocks were not analyzed. Reaction times were summarized by trimmed means; error trials were not included in response time analysis. In the mask identification task, response times shorter than 100 ms or longer than 1199 ms were eliminated as outliers (0.10 % in single task; 0.94 % in triple task).[2] For averaged data, repeated-measures analysis of variance (ANOVA) was performed with factors of consistency (*C*), SOA (*S*), and task type (*T*). Error rates and response accuracy were arcsine-transformed to meet ANOVA requirements. For clarity, all results are reported with Huynh-Feldt-corrected *p* values but the original degrees of freedom, and effects are specified by subscripts to the *F*-values (e.g., $F_{CxS}$ for the interaction of consistency and SOA). While these models are designed to generalize to new participants, data were also analyzed within each participant. For response times and PAS ratings, this was done by ANOVA of individual trials; for accuracy data and error rates, the ANOVA was realized via SPSS's logistic regression module. Note that those models generalize to further trials from a given participant. Throughout the paper, we report all ANOVA effects significant at $p \leq .05$, so that unreported effects are always nonsignificant, with the understanding that *p* values between .01 and .05 should be regarded with caution. We may mention *p* values between .05 and .10 if important to the argument.

In multi-factor repeated-measures designs, statistical power can be calculated if all effect sizes can be predicted along with their respective error variances. In practice, however, too many terms are unknown for a meaningful power analysis. Because the number of trials per participant and condition is about as important for power as the number of participants (Arend & Schäfer, 2019; Baker et al., 2020; Smith & Little, 2018), we control *measurement precision* at the level of individual participants in single tasks and stimulus conditions (Biafora & Schmidt, 2019). For each task, we calculate precision as $s/\sqrt{r}$ (Eisenhart, 1962), where *s* is a single participant's standard deviation in a given cell of the 2x4-design (Consistency x SOA) and *r* is the number of repeated measures in each cell and subject. With *r* = 180 and 270 in the single and triple task conditions, respectively, we expect a precision of about 4.5 ms and 3.7 ms in response times (assuming individual SDs around 60 ms), and at most 3.7 and 3.0 percentage points in accuracy scores (assuming the theoretical maximum SD of .5). Precision thus exceeds our previous recommendations for response priming studies ($r = 60$, F. Schmidt et al., 2011).

*Results*

*Response times and error rates in mask identification.* In the mask ID task, we expected response priming effects in response times and error rates. Because response priming is generated by a response conflict that is aggravated when the prime has more time to impact the response, priming effects in both measures should increase with prime-mask SOA, with response errors predominantly occurring in inconsistent trials at long SOAs (Panis & Schmidt, 2016; F. Schmidt et al., 2011; Vorberg et al., 2003). Due to our assumption that triple tasks cause higher cognitive load and divided attention, we expected longer response times, smaller priming effects, and more errors under triple-task conditions compared to the single task.

Fig. 2 shows response times and error rates in individual participants and the results of within-subject ANOVAs (see Appendix C for test statistics and *p* values). In single tasks, most participants show regular response priming effects (faster responses in consistent than in inconsistent trials) that tend to increase with SOA. Priming effects also occur in the triple task, but response times are extremely delayed (in Participant 5, by about 300 ms) and clearly increase with SOA in only one participant (2). In some participants, priming effects are larger in the triple than in the single task. Four participants show very large priming effects at the shortest SOA, and for Participants 3 and 6 this SOA even yields the largest priming effects.

Despite those differences between participants, the average data provide a clear pattern of effects (Fig. 3). ANOVA with factors of consistency (*C*), SOA (*S*), and task type (*T*) showed that responses were slower (by a massive 162 ms) in the triple task than in the single task, $F_T (1, 7) = 37.57$, $p < .001$. Overall, responses were faster for consistent than for inconsistent trials, $F_C (1, 7) = 21.12$, $p = .002$. As expected, this priming effect increased over SOA for single tasks, but surprisingly decreased with SOA in the triple task (Fig. A1, center), resulting in a three-way interaction, $F_{T \times C \times S} (3, 21) = 5.97$, $p = .004$. Separate ANOVAs for the two task types confirmed significant priming effects in both single and triple tasks, $F_C (1, 7) = 23.45$ and 9.20, $p = .002$ and .019, respectively, and an increase in priming with SOA for single tasks, $F_{C \times S} (3, 21) = 6.34$, $p = .003$. The decrease in priming in the triple task, which is mostly due to Participants 3 and 6 at the shortest SOA, was not significant.

The analogous analysis of the error rates revealed significant priming effects for single as well as triple tasks. Error rates were higher in inconsistent than in consistent trials, $F_C (1, 7) = 6.93$, $p = .034$, and this priming effect increased with SOA, $F_{C \times S} (3, 21) = 3.81$, $p = .025$. Analyses performed separately for the two task types revealed a main effect of consistency that was significant in the triple task, $F_C(1, 7) = 6.14$, $p = .042$, but not in the single task, $F_C(1, 7) = 3.69$, $p = .096$. An interaction with SOA was only observable in the single task, $F_{C \times S} (3, 21) = 3.90$, $p = .026$. Generally, the pattern of errors was in agreement with the pattern of response times.

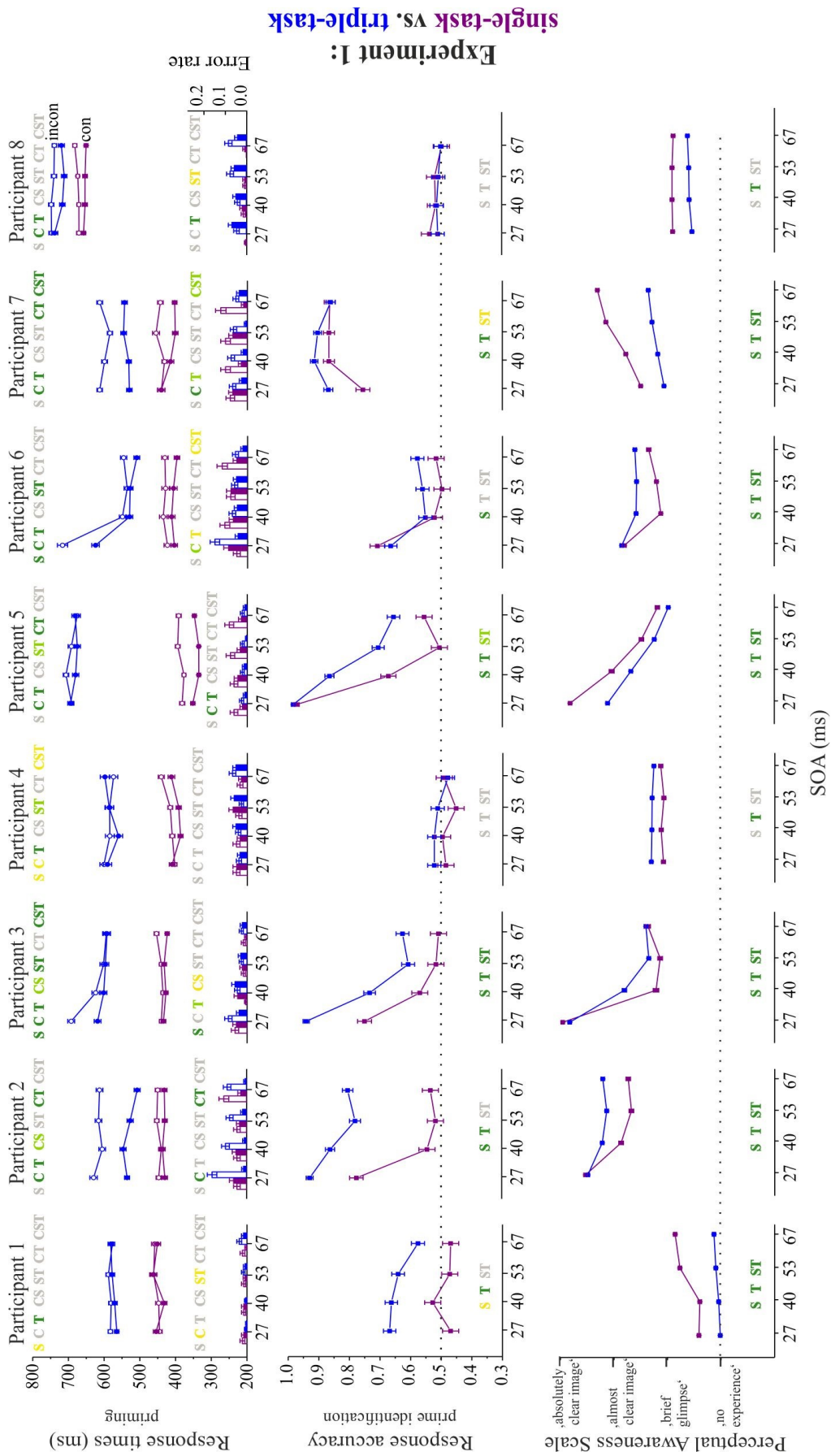

*Fig. 2.* Experiment 1. Individual results of the mask identification task (upper row, response times and error rates), the prime identification task (middle row, response accuracies), and of perceptual awareness ratings (bottom row, perceptual awareness scale) in all eight participants. Letters denote main effects and interactions in within-participant ANOVAs. Factors are coded S (SOA), C (Consistency), and T (Task). Grey: $p > .05$; yellow: $p < .05$; light green: $p < .01$; dark green: $p < .001$.

Rapid-Chase Theory predicts fast errors in inconsistent trials because those responses are produced by feedforward processing of the inconsistent prime. Therefore, incorrect responses to the mask should be as fast as the fastest correct responses. We checked this by applying an ANOVA model to the trimmed responses, treating response accuracy and SOA as fixed factors, and including participants as a random factor to control for repeated measures. Because of the low and unbalanced error rates, this test had to be carried out on the level of single trials, drawn from all participants. We applied this model to the inconsistent trials in single and triple tasks. Because this analysis uses a dependent variable as a factor it is quite unbalanced, but we are only interested in the main effects of response accuracy. For inconsistent trials in single tasks, error responses were 85 ms faster than correct responses $F(1, 27.76) = 120.74$, $p < .001$, while in the triple task the difference was less than 2 ms, $F(1, 9.82) = 0.02$, $p = .903$. Task types also differed in whether the error RTs in inconsistent trials were time-locked to the prime or to the mask, which can be tested by regressing mask-locked and prime-locked response times to the prime-mask SOA. These two slopes would be -1 and 0 if errors were perfectly time-locked to the prime, and 0 and 1 if errors were perfectly time-locked to the masks. In the single task, the slopes were -0.71 ($p = .021$) and 0.29 ($p = .337$), which indicates that responses are (imperfectly) locked to the prime rather than the mask. In the triple task, however, the slopes were -1.95 ($p < .001$) and -0.95 ($p = .040$), which are inconsistent with both prime-locking and mask-locking.

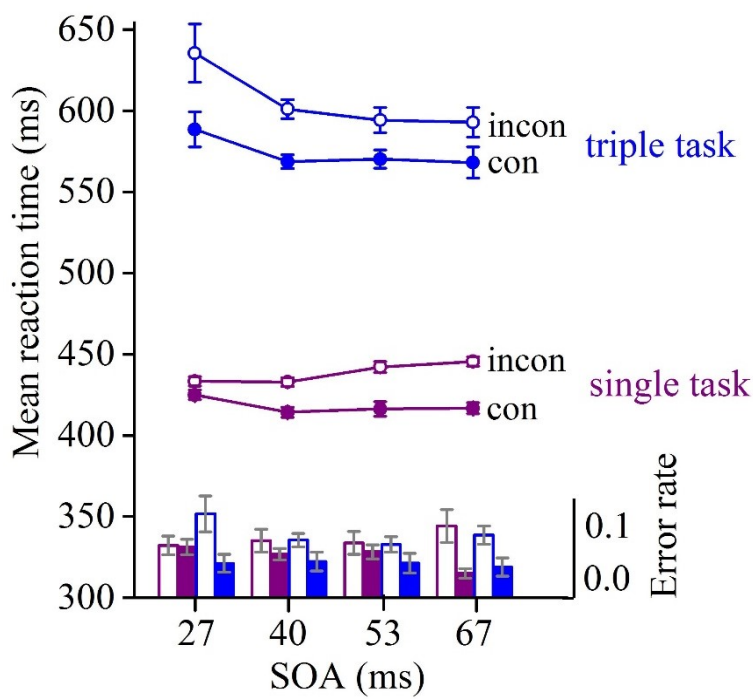

*Fig. 3.* Experiment 1. Mean reaction times and error rates in the mask identification task. Here and in all further plots where participants are averaged, standard errors of the mean are corrected for intersubject variance (Cousineau, 2005). The correction was performed separately for single and triple tasks.

*Accuracy in prime discrimination and PAS ratings.* We used the percentage of correct responses as an objective measure of prime discriminability. Under complete masking, prime discriminability should be at chance level (50 %).

Participants differed markedly in their masking functions (Fig. 2). Except for Participants 4 and 8, all participants performed significantly better in the prime identification task under the triple-task condition, contrary to our expectations. This is illustrated best in

Participant 1, who performed at chance under single-task conditions but reached an accuracy of 63.7 % in the context of a triple task. This finding is a surprise because we expected prime ID performance to be impaired under the triple-task condition due to higher cognitive load and divided attention. The performance advantage in the triple-task condition is most likely an effect of training, because the triple-task condition was always performed after the single-task conditions. We had used this procedure because we expected the challenging triple-task condition to benefit from single-task training, so that our hypothesis of a triple-task disadvantage could be tested conservatively. When performance is traced over sessions, most participants show a smooth learning curve in prime discrimination performance across the single-task session and the subsequent triple-task sessions, consistent with a gradual effect of training.

Otherwise, the participants' individual data patterns differ strongly and even qualitatively. Four participants (2, 3, 5, and 6) show type-B masking, with discrimination performance decreasing with SOA, sometimes strongly. Two more participants (4 and 8) perform at chance level throughout; both show clear evidence of response priming. In contrast, Participant 7 consistently performs around 80 to 90 % accuracy. Finally, Participant 1 performs at chance in the single task but around 65 % correct in the triple task. Because of these qualitative differences between observers, it is not advisable to average them to perform the usual omnibus analysis of variance. Nevertheless, interested readers can find plots and analyses in Appendix A.

Another surprise is that prime discrimination accuracy is often dissociated from the PAS ratings. For example, Participant 8 gives higher PAS ratings in the single than in the triple task, yet his or her discrimination accuracy is at chance in both cases. Participants 4 and 8 both perform objectively at chance but consistently give PAS ratings indicating a "brief glimpse". In Participants 1, 5, and 7, prime discrimination is more accurate in the triple than in the single task, but PAS ratings are lower. This indicates that both tasks do not measure awareness in the same way, or more explicitly, that each of them provides different kinds of information. Again, the qualitative differences between individual PAS functions make it inadvisable to average the data to perform an omnibus ANOVA, but interested readers can find one in Appendix A.

*Discussion*

Experiment 1 shows that the change from single-task to triple-task conditions has dramatic consequences for the pattern of dissociations between direct and indirect tasks. When prime identification performance was measured in a single task, participants scored either near chance or showed a type-B masking function that decreased with SOA. At the same time, these participants show increasing priming effects and thus opposite effects of SOA on priming and masking (double dissociation; Mattler, 2003; Vorberg et al., 2003). In sum, we observed double dissociation patterns in four of the eight participants, simple dissociations in two participants, and no dissociations in another two.

If the same measures are obtained under triple-task conditions, the pattern of results changes. First, there are large main effects: response times are 162 ms slower than under single-task conditions, and prime ID performance is markedly higher. The smooth learning curve in prime ID performance across the single-task session and the subsequent triple-task sessions suggests that participants tried to optimize this performance to the detriment of the mask identification task, which suffers greatly from the increase in cognitive load and divided attention. Second, all double dissociations between priming effects and prime ID performance are lost: on average, priming effects would even *decrease* with SOA, together with prime identification accuracies (even though this is due to only two participants in only one SOA). While this loss of dissociation is in line with Peremen and Lamy's (2014) findings, we do not believe that it reveals an artifact created by the single-task measurement, but rather a disruption of the response activation process by the triple task. We suspect that the triple task slows responses to the mask so much that the effects are no longer based on simple feedforward processing: it may instead occur out of a memory representation and even be subject to response inhibition. The absence of the characteristic early and rapid errors from the triple-task condition and the absence of time-locking of errors to the onset of the prime is another indicator that the prime has lost the power to drive motor responses to completion.

Finally, the patterns of prime ID performance and PAS ratings are inconsistent. On average, prime ID performance decreases with SOA and is higher in the triple task, while PAS ratings are invariant across all those conditions. Upon closer analysis, however, it is evident that some participants show higher PAS ratings in the triple task while others show lower ones, and that ratings can increase, decrease, or stay invariant across SOAs in individual observers. In addition, participants who perform at chance in the objective task do not give zero ratings in the subjective task. Because neither the PAS ratings nor the prime ID accuracies are particularly noisy, we have to assume that the two tasks capture different stimulus aspects and thus differ in criterion content (Kahneman, 1968).

**Experiment 2**

In the first experiment, we compared speeded mask ID, unspeeded prime ID, and PAS ratings under single- and triple-task conditions. We found that the triple task condition did not impair prime ID performance, which steadily increased across sessions, but greatly impaired response priming effects, resulting in strongly delayed responses, decreasing instead of increasing priming effects, and a loss of fast prime-locked errors in inconsistent trials. Whereas both single and double dissociations were observed under single-task conditions, most of these dissociations were lost in the triple task. In Experiment 2, we investigate whether similar impairments arise under dual-task conditions where only response priming and prime discrimination are measured on the same trial.

*Methods*

*Participants.* Eight volunteers, mostly students from the University of Kaiserslautern (4 male; age range 22-25 years) took part in four 1-hour sessions. None of them had participated in Experiment 1, and all of them were naïve to the purpose of the study. Participants were recruited in the course of bachelor theses, which is why attendees did not receive payment or course credit. Their vision was normal or corrected to normal. Each of the participants gave informed consent and was treated according to the ethical guidelines of the APA. All of them were debriefed and received an explanation of the experiment after the final session.

*Apparatus, stimuli and procedure.* The equipment was the same as in Experiment 1. We also used the same stimuli and procedures, but with the exception that participants performed only two tasks (mID, pID) within the same trial (*dual-task* condition), or in different sessions (*single-task* condition). This time, no subjective awareness rating was employed.

All participants performed 31 blocks of 48 trials in the single-task condition and of 36 trials in the dual-task condition to obtain session times of equal duration. Due to a programming mistake, the final block stopped after 28 trials in the dual task. For both conditions, the first block was always a practice block. Each participant performed one session of mask ID, followed by one session of prime ID, followed by two sessions of the dual task. All stimulus combinations were presented equiprobably and pseudo-randomly in each block.

*Data treatment and statistical methods.* Data treatment proceeded as in Experiment 1. In the mask ID task, 0.05 and 0.16 % of trials were discarded as outliers in the single and dual task, respectively. Repeated-measures analysis of variance (ANOVA) was performed with factors of consistency ($C$), SOA ($S$), and task-type ($T$) on response times and arcsine-transformed error rates. Measurement precision ($s/\sqrt{r}$, with $r = 180$ and 270 in the single and double task conditions) was the same as in Experiment 1.

*Results*

*Response times and error rates in mask identification.* Fig. 4 shows response times and error rates in individual participants and the results of within-subject ANOVAs (see Appendix C for test statistics and *p* values). In single tasks, most participants show regular response priming effects (faster responses in consistent than in inconsistent trials) that tend to increase with SOA. Response times in the dual task are strongly delayed, but not nearly as much as in the triple task of Experiment 1. Priming effects are generally larger in the dual than in the single task; importantly, they generally increase with SOA (except for Participant 5, who shows the largest priming effect at the shortest SOA).

Even though participants differ markedly in the magnitude of their priming effects, the average data provide a clear pattern of effects (Fig. 5). Responses were faster for consistent than for inconsistent trials, $F_C (1, 7) = 21.87$, $p = .002$, and this priming effect increased over SOA, $F_{CxS} (3, 21) = 4.57$, $p = .020$. Response times were on average 76 ms faster in the single than in the dual-task condition, $F_T (1, 7) = 15.84$, $p = .005$, while priming effects were larger in the dual task, $F_{TxC} (1, 7) = 7.10$, $p = .032$. ANOVAs performed separately for single and dual tasks revealed significant main effects of consistency, $F_C(1, 7) = 16.94$ and 17.20, both $p = .004$, and a significant main effect of SOA for the single task, $F_S(3, 21) = 3.91$, $p = .043$, but no significant interaction effects in either task.

We also performed an analogous ANOVA of the arcsine-transformed error rates, showing that more errors occurred in inconsistent trials, $F_C(1, 7) = 11.96$, $p = .011$, and that this priming effect increased with SOA, $F_{C\times S}(3, 21) = 10.70$, $p < .001$. Overall, error rate tended to increase with SOA, $F_S(3, 21) = 3.14$, but not significantly, $p = .054$. The remaining effects were nonsignificant. ANOVAs performed separately for single and dual tasks revealed a significant main effect of consistency only for the dual task, $F_C(1, 7) = 12.20$, $p = .010$, but not for the single task, $F_C(1, 7) = 4.30$, $p = .077$. Priming effects increased with SOA both in the single task, $F_{C\times S}(3, 21) = 4.28$, $p = .021$, and in the dual task, $F_{C\times S}(3, 21) = 9.29$, $p = .001$. Error rates increased with SOA in the dual task only, $F_S(3, 21) = 4.06$, $p = .020$.

We checked for fast errors in inconsistent conditions by applying the same ANOVA models as in Experiment 1 to single and dual tasks. For single tasks, error responses in inconsistent trials were 62 ms faster than correct responses, $F(1, 8.12) = 27.21$, $p = .001$, while in the dual task they were only 20 ms faster, $F(1, 8.76) = 2.78$, $p = .131$. We checked for time-locking of these inconsistent error trials using the same regression technique as in Experiment 1. When regressing mask-locked and prime-locked response times to the SOA, the two slopes would be -1 and 0 if errors were perfectly time-locked to the prime, and 0 and 1 if errors were perfectly time-locked to the mask. The slopes were -1.10 ($p = .007$) and -0.10 ($p = .813$) in the single task, and -1.06 ($p = .049$) and -0.06 ($p = .912$) in the dual task, indicating near-perfect time-locking to the prime in both task types.

*Accuracy in prime identification.* Again, participants differed markedly in their masking functions (Fig. 4). Under single-task conditions, Participants 1, 2, 3, and 4 basically performed at chance level at all SOAs when trying to identify the prime but showed huge increases in performance in the dual task. In the remaining participants, dual-task gains are much smaller (Participants 5 and 7) or absent (Participants 6 and 8). Note that this new group of observers generally showed type-A masking functions. Because of these qualitative differences between observers, it is not advisable to average them to perform the usual omnibus analysis of variance. Nevertheless, interested readers can find plots and analyses in Appendix B.

Except for Participant 6, all participants performed the mask ID task slower in the dual task than in single task; in some participants, these dual-task losses were massive. This data pattern is in line with the findings of Experiment 1. Nevertheless, in Participants 3, 4, and 8, priming effects are larger in the dual task. In Participants 3 and 4, this increase in the priming effect coincides with an increase in prime ID accuracy, but in Participant 8 it does not. There is no indication that priming effects might decrease under dual-task conditions (with the possible exception of Participant 5).

*Discussion*

When prime identification performance and priming effects were each measured in a single task, priming effects increased with SOA while prime ID performance also showed a slight (nonsignificant) increase. Individual participants scored either near chance level or showed a type-A masking function that increased with SOA. At the same time, most participants show increasing priming effects. Unfortunately, none of them showed type-B masking, making it impossible to assess whether any double dissociation (increasing priming effects under decreasing prime discrimination performance) can be observed under dual task conditions.[3] Still, note that prime discrimination performance is not able to predict the magnitude of priming. For example, of the two participants showing only little priming one performed near chance level in prime ID (Participant 3) while the other one was almost perfect (Participant 6). Similarly, there are two participants (6, 8) in which masking functions are virtually identical under both task conditions, while the priming effects are strongly different.

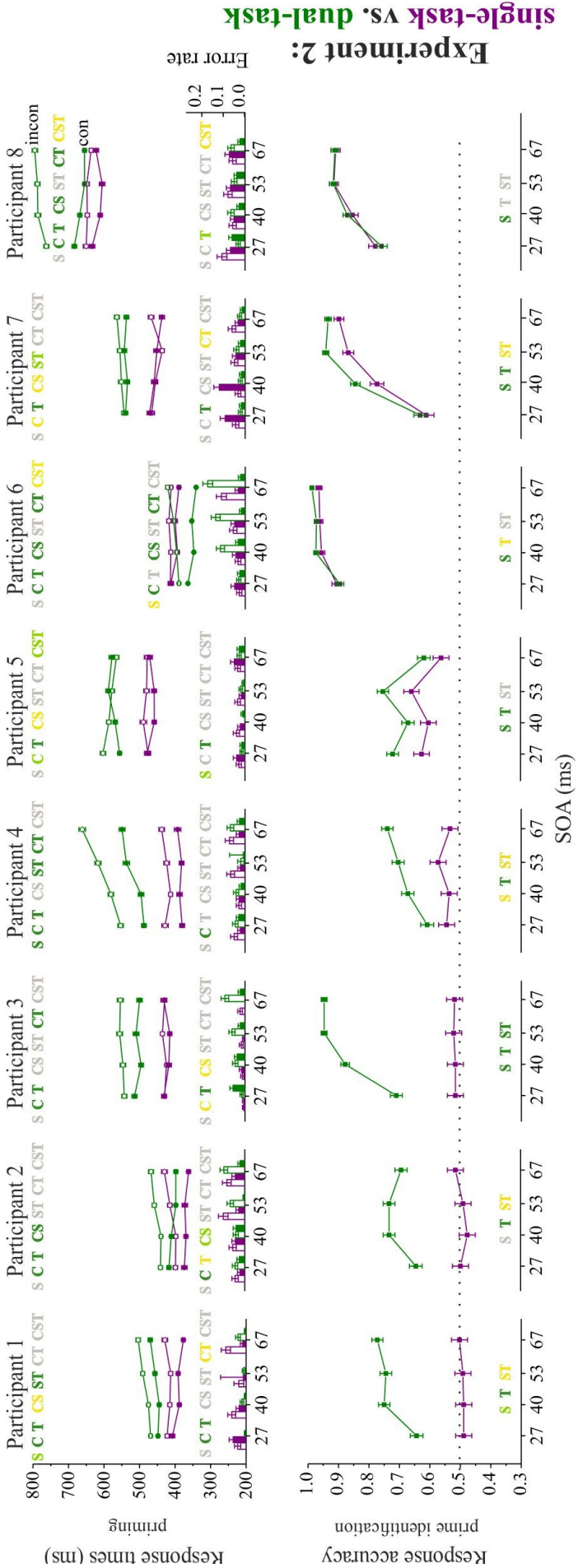

*Fig 4.* Experiment 2. Individual results of the mask identification task (upper row, response times and error rates) and the prime identification task (lower row, response accuracies) in all eight participants. Conventions are the same as in Fig. 2.

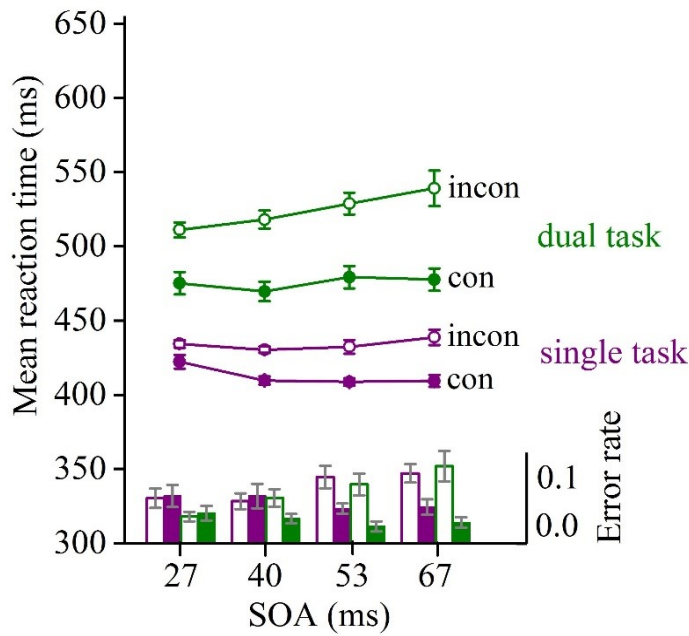

*Fig. 5.* Experiment 2. Mean reaction times and error rates in the mask identification task.

As in Experiment 1, there are systematic changes in the pattern of results when the same measures are obtained under dual-task conditions. Again, accuracy in prime ID is much higher than under single-task conditions and response times are much slower (even though not as slow as in the triple task of Experiment 1) and show larger priming effects. Second, the simple dissociations between priming effects and prime ID performance are lost: those participants who scored at chance in prime ID now score much higher (e.g., Participant 3 improves from chance level to more than 90 % accuracy). This large improvement suggests that participants may be able to use information only available in the dual task.[4]

Perhaps most importantly, the structure of the response priming effects seems altered under dual-task conditions: whereas errors provoked by inconsistent primes are usually fast in single tasks, the dual task not only leads to slower correct responses but also to slower errors, such that the speed difference between correct and incorrect responses is no longer significant. On the other hand, error responses remain time-locked to prime onset. According to Rapid-Chase Theory (T. Schmidt et al., 2006, 2011; see also T. Schmidt, 2014), fast errors occur in inconsistent trials due to the feedforward response activation from the inconsistent prime. Our conclusion is that even though the dual-task situation seems less disruptive for response priming than the triple-task situation, pure feedforward processes are harder to guarantee under dual tasks than under single tasks.

**General Discussion**

In this paper, we compared response priming effects, objective prime discrimination performance, and subjective PAS ratings in single, dual, and triple tasks to address the following set of questions. First, are objective and subjective measures of awareness equally sensitive to changes in stimulus conditions and equally valid for measuring awareness of the critical prime feature? Second, do multitasks lead to qualitative changes in any of the measures, which threaten their validity? Third, do single and multiple tasks both provide the necessary conditions for obtaining a double dissociation?

*A multitask advantage in prime discrimination?* In both Experiments 1 and 2, participants showed better performance in prime discrimination in multitasks than in single tasks. This was most likely caused by our decision to perform the multitask sessions after the single-task sessions to give the multitasks the benefit of optimal practice – we expected prime discrimination to be more difficult under multitask conditions because observers would be forced to divide their attention across stimuli and features, and wanted the triple task to benefit from the additional training to avoid underestimating its sensitivity. Indeed, prime discrimination performance increases rather steadily across all blocks and sessions – in most participants, it forms a relatively smooth power function (however, there is a little upstep in performance between single- and dual-task sessions in Experiment 2).

To clarify the issue, we repeated Experiment 2 with alternating single-task and dual-task sessions, and a new group of participants so that both tasks received comparable practice. We found priming effects that were relatively small but increased with SOA in both task types. Responses were slower in the dual task, and there was a loss of fast errors compared to the single task. Prime discrimination performance was similar in both tasks, supporting our conclusion that the multitask benefits observed here are the result of differential training.

*Comparing objective and subjective measures.* Experiment 1 compared objective prime discrimination performance and subjective PAS ratings. On average, performance in prime ID is markedly higher in the triple than in the single task and decreases with SOA, while PAS ratings are relatively more constant over conditions at an average level roughly corresponding to a "brief glimpse" rating. At first glance, this pattern seems inconsistent with the claim that PAS ratings were as sensitive as objective measures of performance (Avneon & Lamy, 2018; Lamy et al., 2015, 2017; Peremen & Lamy, 2014). However, it is necessary to consider the data patterns of individual participants because masking effects differ systematically between observers and masking functions should not be averaged. Of eight observers, four showed a decrease in prime ID performance with SOA, in some cases over the full range of the scale. Two more observers performed at chance level throughout, one high-performing observer showed a slight increase, and one low-performing observer was at chance in the single task but slightly better in the triple task. If we compare the PAS ratings with objective discrimination performance, we see that increases, decreases, and low constant values in prime ID and PAS ratings co-occur in the same observers, which means that both measures do have some sensitivity to experimental conditions. However, visibility ratings of "clear experience" and "no experience" rarely occur even if observers perform perfectly or near chance in the objective measure, and objective chance performance seems to be associated with ratings of "brief glimpse" rather than "no experience". Compared to objective prime discrimination effects, PAS effects seem compressed towards medium values, suggesting that the PAS scale is not good at capturing extreme values of discriminability.

There are also some striking discrepancies between the measures: three observers (1, 5, and 7) perform better under triple-task than under single-task conditions in the objective measure (likely as a result of training effects), but show the reverse order in the subjective measure. This reversal implies that the objective prime ID task and the subjective awareness rating cannot register exactly the same information. Observers may use cues to identify the prime that the PAS simply does not ask about – in other words, there would be some mismatch between the wording of the PAS labels and the criterion content (Kahneman, 1968) observers actually use when identifying the prime. Other examples would be percepts like rotation or expansion in the prime-mask sequence that can help to infer the prime (Albrecht & Mattler, 2012; Koster, Mattler, & Albrecht, 2020; T. Schmidt, 2000). If observers use the PAS labels literally, they may still give truthfully low visibility ratings because the objective performance does not derive from prime visibility in the first place but from such auxiliary cues. For instance, an observer may have a clear impression of rotation in the prime-mask sequence despite getting only a "brief glimpse" of the prime. Similarly, an observer may truthfully report a "brief glimpse" indicating that the prime was detected, even if it could not be discriminated.

Is the PAS a valid measure of prime visibility, then? This is a question of whether or not the scale labels match the criterion content actually used by the observer in the task at hand. Ramsøy and Overgaard (2004) introduced their scale in an experiment where the critical stimulus was one of three shapes appearing in one of three colors and at one of three positions. The scale was applied separately to each of those stimulus dimensions, which means that the feature of interest was specified first and then the scale was applied specifically to it (e.g., "clear impression of the location", "brief glimpse of the shape", "brief glimpse of the color"). Such specification increases the likelihood that the scale labels are applied to the actual criterion content (or maybe that the observer deliberately uses the criterion content that the scale intends to measure). But like the vast majority of users, we applied the scale without specifying the aspect of stimulus experience to apply it to. As it stands, the PAS asks rather unspecifically about "the stimulus", leaving it entirely to the participant to fill this chiffre with perceptual meaning. But if we want to use the scale to evaluate the possibility of unconscious response priming, we need it to focus on the *critical feature* that generates the priming effect – the distinction between square and diamond primes. It is this critical feature that would establish the dissociation between tasks, and any other measure would lead to a mismatch between direct and indirect tasks (*D-I mismatch*, T. Schmidt & Vorberg, 2006). These considerations imply that the PAS can rarely just be applied as it is: It needs to be combined with a clear instruction which aspect of subjective experience is to be rated, as Ramsøy and Overgaard (2004) did in their original study.

*The structure of priming under single, dual, and triple tasks.* Experiment 1 clearly shows that the triple-task situation interferes with response priming. The triple task leads to drastically slower responses and decreasing instead of increasing priming effects over the course of the prime-mask SOA. The double dissociation observed under single-task conditions (priming effects increase with SOA while prime ID performance decreases) is therefore lost under triple-task conditions. This is consistent with the findings of Peremen and Lamy (2014), who reported u-shaped priming functions in tandem with u-shaped masking functions under triple-task conditions and conclude that "[*d*]*ifferences in attention rather than in conscious perception may therefore account for the dissociated time courses*" in single tasks (Peremen & Lamy, 2014, p. 23). This sounds as if the single-task measurement created an artifact in attentional employment: the authors seem to expect that priming effects should increase with SOA whenever attention is fully directed to the mask, but decrease with SOA when attention is shared with the prime. This is inconsistent with several studies that investigated the effect of different aspects of attention on the time-course of response priming. When primes and masks appear at an unattended location, priming effects are smaller and responses are slower compared to an attended location, but they still increase with SOA (T. Schmidt & Seydell, 2008; cf. Sumner, Tsai, Yu, & Nachev, 2006). Attention modulates the entire time-course of the motor response, including response onset, maximum effector velocity, and completion time. The same modulation of priming effects occurs when a non-positional feature (e.g., a color or a shape) is cued, even if the prime remains indiscriminable (F. Schmidt & T. Schmidt, 2010). In these experiments, attention was always manipulated simultaneously for prime and target, which should have an effect similar to splitting attention between both stimuli. Szumska, Baran, Pinkas, and Van der Lubbe (2019) compared prime and mask identification tasks and found that response activation by the prime is much smaller in pID than in mID tasks, but such an effect should as well lead to priming effects that are smaller in magnitude, not reversed. Finally, directly varying prime and target contrast independently modulates priming effects and overall response time but does not reverse the time-course of priming (Vath & Schmidt, 2007).

In our opinion, what happens in the triple task cannot be described by a simple reallocation of visual attention. Rather, it seems that the need to divide attention causes massive cognitive costs (Pashler, 1989, 1994b) associated with higher working memory load and cognitive control. The triple task requires participants to attend to three different aspects of the trial: the *shape of the mask*, the *shape of the prime*, and the *subjective impression of prime clarity*. Whereas speeded responses to the mask would normally be automatized, the two direct measures require consideration of multiple sources of information. Therefore, all relevant information has to be held in working memory while a sequence of three responses is scheduled. In this process, most participants prioritize the prime-related responses, as suggested by the fact that prime discrimination performance does not suffer from the triple task. Instead, the prioritization seems to be entirely to the detriment of the speeded mask ID task. Experiment 1 clearly shows that the triple task leads to a structural change in priming effects: As compared to the single task, responses to the mask are slowed by more than 160 ms, the priming effect is decreasing with SOA instead of increasing, fast errors disappear, and error RTs are no longer time-locked to the prime (or any other stimulus). Fast errors are one of the indicators that response priming is feedforward: Because Rapid-Chase Theory assumes that the earliest responses are controlled exclusively by the prime, it predicts that prime-provoked errors should be as fast as the fastest correct responses (T. Schmidt, 2014).

The slowing of responses also explains the otherwise puzzling finding that priming effects decrease with SOA under triple-task conditions. If the prime-mask SOA is prolonged or if responses occur very late, they are often subject to response inhibition. Response inhibition can not only reduce the magnitude of the priming effect, but even reverse it such that responses are faster in inconsistent than in consistent trials (*negative compatibility effect or NCE*; Eimer & Schlaghecken, 1998, 2003; see Sumner, 2008, for a review). T. Schmidt, Hauch, and F. Schmidt (2015) show that the NCE in pointing movements results from inhibition of the primed response and activation (or disinhibition) of the opposite response, as shown by late errors that start in a spatial direction 180° opposite to both the prime and target direction (cf. Boy, Husain, & Sumner, 2010; Jaśkowski, & Przekoracka-Krawczyk, 2005). Panis and Schmidt (2016) analyze the NCE in keypress responses and find that it is associated with late errors in consistent trials starting about 320 ms after mask onset (in a sequence of prime, mask, and target). Importantly, the NCE is most pronounced in slow responses (Ocampo & Finkbeiner, 2013), and a reanalysis of the priming effects in the triple task of Experiment 1 reveals that the later the decile of the response time distribution (calculated separately for participants and conditions), the more pronounced the decrease in priming with SOA. In our opinion, inhibition of the primed response is the most likely source of the altered time-course of priming in the triple task. In sum, it seems that the triple-task situation does not prevent priming effects altogether, but it leads to a structural change in priming that is no longer consistent with sequential feedforward response activation by primes and masks. Instead, our data suggest that priming now occurs out of a memory representation under high cognitive load and is subject to response inhibition. But if feedforward response activation is independent of awareness because it works without recurrent processing (Lamme & Roelfsema, 2000), disrupting the feedforward mechanism of priming would eliminate the necessary conditions for priming without awareness.

If the triple task is so disruptive for response priming, what about the dual task? Overall, the dual task is much more benign to the microstructure of response priming. Priming effects continue to increase with SOA, and error responses in inconsistent trials are time-locked to the prime. In some participants, priming effects were even larger in the dual task. However, the dual task still slows responses to the prime by more than 70 ms, and error responses are no longer significantly faster than correct responses, indicating that the feedforward nature of response priming is at least under strain from the increased demands in memory load and cognitive control. These findings indicate that even dual tasks should be used with caution, and that their robustness should be further examined.

# Appendix

Individual differences in the time-course of visual masking were too large to base our conclusions on averaged data. However, we expect that critically-minded readers may wish to inspect the averaged masking and PAS functions anyway together with the respective ANOVAs.

### A: Experiment 1, averaged accuracy and PAS functions

*Accuracy.* Averaged across participants, ANOVA showed a significant main effect of task type, $F_T(1, 7) = 9.91$, $p = .016$, and a less reliable main effect of SOA, $F_S(3, 21) = 4.82$, $p = .056$ (Fig. A1). No interaction of these two factors was found, $F_{TxS}(3, 21) = 1.22$, $p = .326$. Simple tests showed that the SOA effect was significant in the triple task but not in the single task, $F_S(3, 21) = 5.34$ and $3.81$, $p = .033$ and $.088$, respectively. Note that no single participant's response pattern closely resembled this average pattern (Fig. 2).

*PAS ratings.* We expected the PAS ratings to be qualitatively similar to the masking functions derived from the pID task, and we had similar expectations for the outcome. However, ANOVA showed no significant effects at all (Fig. A1), neither a difference between single and triple task, $F_T(1, 7) = 0.72$, $p = .424$, nor a main effect of SOA, $F_S(3, 21) = 2.09$, $p = .184$, nor an interaction, $F_{TxS}(3, 21) = 1.80$, $p = .200$. Note that this apparent null result is highly misleading; it belies the fact that individual PAS functions were highly reliable but differed qualitatively across observers (Fig. 2).

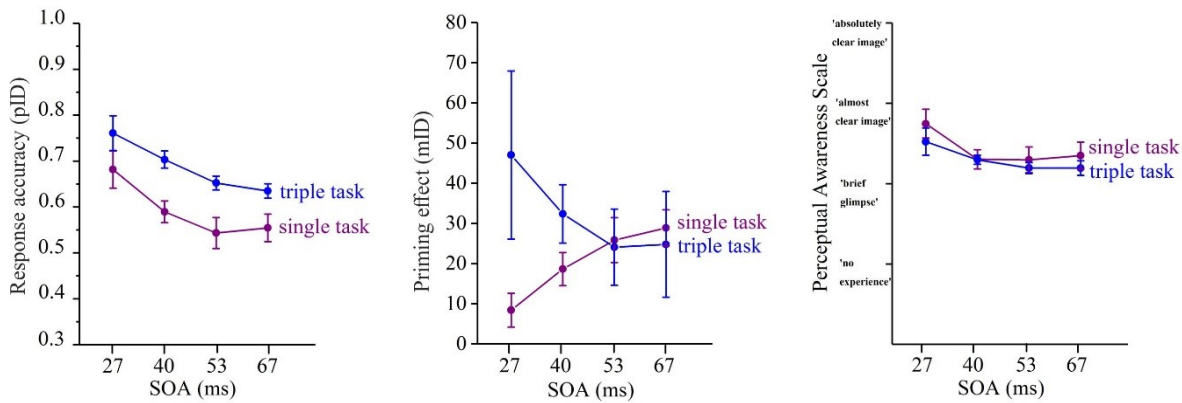

*Fig. A1.* Experiment 1. *Left:* Average response accuracy in the prime identification task. *Center:* Average response-time priming effects ($RT_{incon} - RT_{con}$) in the mask identification task. *Right:* Average ratings on the Perceptual Awareness Scale.

### B: Experiment 2, averaged accuracy functions

Averaged across participants, ANOVA showed a significant main effect of task type, $F_T(1, 7) = 9.91$, $p = .016$, a main effect of SOA, $F_S(3, 21) = 5.89$, $p = .036$, and a significant interaction indicating that the increase with SOA was steeper in the dual than in the single task, $F_{TxS}(3, 21) = 3.81$, $p = .028$. Simple tests showed that the SOA effect was significant in the dual task but not in the single task, $F_S(3, 21) = 7.25$ and $2.40$, $p = .031$ and $.156$, respectively. As in Experiment 1, no single participant's response pattern closely resembled this average pattern (Fig. 4).

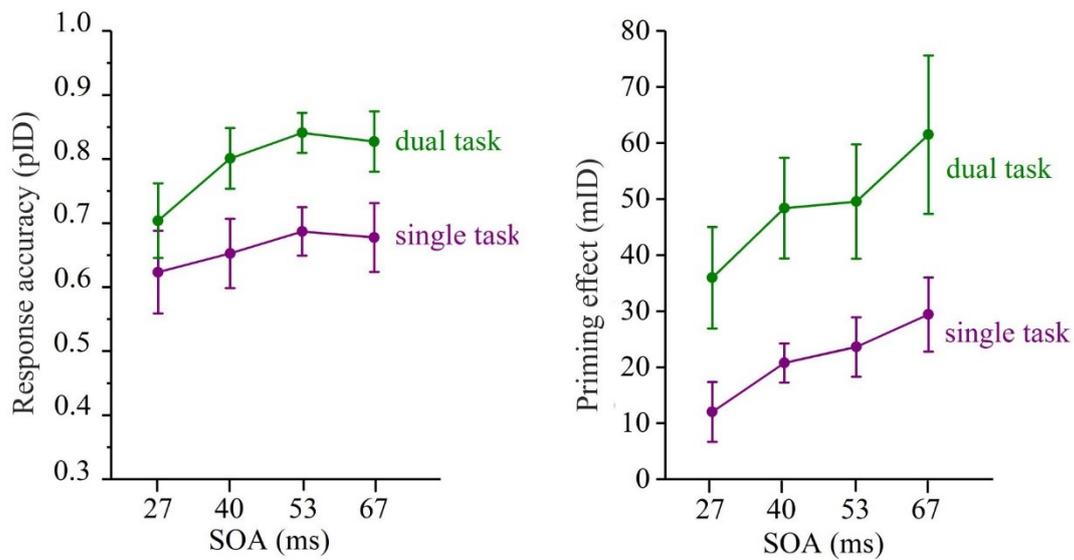

Fig. A2. Experiment 2. *Left:* Average response accuracy in the prime identification task. *Right:* Average response-time priming effects in the mask identification task.

**C: Results for individual participants in Experiment 1.**

Table A1 holds the test statistics and *p* values for the individual main effects and interactions reported in graphical form in Fig. 2.

Table A1: Test statistics, degrees of freedom, and *p* values for the main effects and interactions reported in Fig. 2. The large number of denominator degrees of freedom is determined by the number of trials per participant. Note that these analyses aim to generalize to new trials from the same participant rather than to new participants.

### P1

**Response times:**
- $F_S$ (3, 3510) = 2.672, $p = .046$
- $F_C$ (1, 3510) = 3.088, $p = .079$
- $F_T$ (1, 3510) = 1206.399, $p < .001$
- $F_{CS}$ (3, 3510) = 0.464, $p = .707$
- $F_{ST}$ (3, 3510) = 0.835, $p = .474$
- $F_{CT}$ (1, 3510) = 0.660, $p = .417$
- $F_{CST}$ (3, 3510) = 0.997, $p = .393$

**Errors:**
- $\chi^2_S$ (3, N=3597) = 4.477, $p = .214$
- $\chi^2_C$ (1, N=3597) = 5.709, $p = .017$
- $\chi^2_T$ (1, N=3597) = 3.251, $p = .071$
- $\chi^2_{CS}$ (3, N=3597) = 1.337, $p = .720$
- $\chi^2_{ST}$ (3, N=3597) = 11.422, $p = .010$
- $\chi^2_{CT}$ (1, N=3597) = 1.065, $p = .302$
- $\chi^2_{CST}$ (3, N=3597) = 2.567, $p = .463$

**Response Accuracy in pID:**
- $\chi^2_S$ (3, N=3600) = 9.996, $p = .019$
- $\chi^2_T$ (1, N=3600) = 82.437, $p < .001$
- $\chi^2_{ST}$ (3, N=3600) = 4.494, $p = .213$

**Ratings on PAS:**
- $F_S$ (3, 3592) = 141.770, $p < .001$
- $F_T$ (1, 3592) = 1945.052, $p < .001$
- $F_{ST}$ (3, 3592) = 59.764, $p < .001$

### P2

**Response times:**
- $F_S$ (3, 3333) = 1.284, $p = .278$
- $F_C$ (1, 3333) = 147.209, $p < .001$
- $F_T$ (1, 3333) = 1070.208, $p < .001$
- $F_{CS}$ (3, 3333) = 3.521, $p = .014$
- $F_{ST}$ (3, 3333) = 2.308, $p = .075$
- $F_{CT}$ (1, 3333) = 58.699, $p < .001$
- $F_{CST}$ (3, 3333) = 0.477, $p = .699$

**Errors:**
- $\chi^2_S$ (3, N=3585) = 1.810, $p = .613$
- $\chi^2_C$ (1, N=3585) = 48.499, $p < .001$
- $\chi^2_T$ (1, N=3585) = 2.092, $p = .148$
- $\chi^2_{CS}$ (3, N=3585) = 4.604, $p = .203$
- $\chi^2_{ST}$ (3, N=3585) = 2.293, $p = .514$
- $\chi^2_{CT}$ (1, N=3585) = 25.494, $p < .001$
- $\chi^2_{CST}$ (3, N=3585) = 3.732, $p = .292$

**Response Accuracy in pID:**
- $\chi^2_S$ (3, N=3600) = 120.791, $p < .001$
- $\chi^2_T$ (1, N=3600) = 272.129, $p < .001$
- $\chi^2_{ST}$ (3, N=3600) = 4.687, $p = .196$

**Ratings on PAS:**
- $F_S$ (3, 3592) = 112.209, $p < .001$
- $F_T$ (1, 3592) = 144.957, $p < .001$
- $F_{ST}$ (3, 3592) = 22.066, $p < .001$

### P3

**Response times:**
- $F_S$ (3, 3376) = 11.922, $p < .001$
- $F_C$ (1, 3376) = 17.958, $p < .001$
- $F_T$ (1, 3376) = 1670.165, $p < .001$
- $F_{CS}$ (3, 3376) = 3.526, $p = .014$
- $F_{ST}$ (3, 3376) = 11.842, $p < .001$
- $F_{CT}$ (1, 3376) = 1.555, $p = .213$
- $F_{CST}$ (3, 3376) = 6.053, $p < .001$

**Errors:**
- $\chi^2_S$ (3, N=3575) = 21.852, $p < .001$
- $\chi^2_C$ (1, N=3575) = 0.066, $p = .798$
- $\chi^2_T$ (1, N=3575) = 7.662, $p = .006$
- $\chi^2_{CS}$ (3, N=3575) = 10.272, $p = .016$
- $\chi^2_{ST}$ (3, N=3575) = 4.237, $p = .237$
- $\chi^2_{CT}$ (1, N=3575) = 0.812, $p = .367$
- $\chi^2_{CST}$ (3, N=3575) = 3.507, $p = .320$

**Response Accuracy in pID:**
- $\chi^2_S$ (3, N=3600) = 256.392, $p < .001$
- $\chi^2_T$ (1, N=3600) = 104.626, $p < .001$
- $\chi^2_{ST}$ (3, N=3600) = 29.652, $p < .001$

**Ratings on PAS:**
- $F_S$ (3, 3592) = 827.639, $p < .001$
- $F_T$ (1, 3592) = 44.493, $p < .001$
- $F_{ST}$ (3, 3592) = 33.832, $p < .001$

### P4

**Response times:**
- $F_S$ (3, 3257) = 3.859, $p = .009$
- $F_C$ (1, 3257) = 3.938, $p = .047$
- $F_T$ (1, 3257) = 1022.469, $p < .001$
- $F_{CS}$ (3, 3257) = 1.035, $p = .376$
- $F_{ST}$ (3, 3257) = 4.337, $p = .005$
- $F_{CT}$ (1, 3257) = 2.558, $p = .110$
- $F_{CST}$ (3, 3257) = 2.626, $p = .049$

**Errors:**
- $\chi^2_S$ (3, N=3527) = 0.113, $p = .990$
- $\chi^2_C$ (1, N=3527) = 0.250, $p = .617$
- $\chi^2_T$ (1, N=3527) = 0.214, $p = .644$
- $\chi^2_{CS}$ (3, N=3527) = 4.857, $p = .183$
- $\chi^2_{ST}$ (3, N=3527) = 5.816, $p = .212$
- $\chi^2_{CT}$ (1, N=3527) = 0.316, $p = .574$
- $\chi^2_{CST}$ (3, N=3527) = 0.892, $p = .827$

**Response Accuracy in pID:**
- $\chi^2_S$ (3, N=3600) = 1.997, $p = .573$
- $\chi^2_T$ (1, N=3600) = 2.760, $p = .097$
- $\chi^2_{ST}$ (3, N=3600) = 2.370, $p = .499$

**Ratings on PAS:**
- $F_S$ (3, 3592) = 0.310, $p = .819$
- $F_T$ (1, 3592) = 89.084, $p < .001$
- $F_{ST}$ (3, 3592) = 1.336, $p = .261$

### P5

**Response times:**
- $F_S$ (3, 3393) = 0.863, $p = .459$
- $F_C$ (1, 3393) = 49.261, $p < .001$
- $F_T$ (1, 3393) = 6951.189, $p < .001$
- $F_{CS}$ (3, 3393) = 2.167, $p = .090$
- $F_{ST}$ (3, 3393) = 4.181, $p = .006$
- $F_{CT}$ (1, 3393) = 18.831, $p < .001$
- $F_{CST}$ (3, 3393) = 2.079, $p = .101$

**Errors:**
- $\chi^2_S$ (3, N=3576) = 1.213, $p = .750$
- $\chi^2_C$ (1, N=3576) = 16.271, $p < .001$
- $\chi^2_T$ (1, N=3576) = 28.213, $p < .001$
- $\chi^2_{CS}$ (3, N=3576) = 2.250, $p = .522$
- $\chi^2_{ST}$ (3, N=3576) = 1.858, $p = .602$
- $\chi^2_{CT}$ (1, N=3576) = 0.030, $p = .863$
- $\chi^2_{CST}$ (3, N=3576) = 0.967, $p = .807$

**Response Accuracy in pID:**
- $\chi^2_S$ (3, N=3600) = 534.715, $p < .001$
- $\chi^2_T$ (1, N=3600) = 27.119, $p < .001$
- $\chi^2_{ST}$ (3, N=3600) = 11.689, $p = .009$

**Ratings on PAS:**
- $F_S$ (3, 3592) = 684.138, $p < .001$
- $F_T$ (1, 3592) = 261.438, $p < .001$
- $F_{ST}$ (3, 3592) = 23.843, $p < .001$

### P6

**Response times:**
- $F_S$ (3, 3256) = 34.231, $p < .001$
- $F_C$ (1, 3256) = 31.522, $p < .001$
- $F_T$ (1, 3256) = 979.584, $p < .001$
- $F_{CS}$ (3, 3256) = 1.632, $p = .180$
- $F_{ST}$ (3, 3256) = 36.264, $p < .001$
- $F_{CT}$ (1, 3256) = 1.760, $p = .185$
- $F_{CST}$ (3, 3256) = 1.849, $p = .136$

**Errors:**
- $\chi^2_S$ (3, N=3572) = 3.591, $p = .309$
- $\chi^2_C$ (1, N=3572) = 10.441, $p = .001$
- $\chi^2_T$ (1, N=3572) = 5.464, $p = .019$
- $\chi^2_{CS}$ (3, N=3572) = 4.440, $p = .218$
- $\chi^2_{ST}$ (3, N=3572) = 6.888, $p = .076$
- $\chi^2_{CT}$ (1, N=3572) = 3.253, $p = .071$
- $\chi^2_{CST}$ (3, N=3572) = 9.723, $p = .021$

**Response Accuracy in pID:**
- $\chi^2_S$ (3, N=3600) = 60.616, $p < .001$
- $\chi^2_T$ (1, N=3600) = 2.272, $p = .132$
- $\chi^2_{ST}$ (3, N=3600) = 6.692, $p = .082$

**Ratings on PAS:**
- $F_S$ (3, 3592) = 95.526, $p < .001$
- $F_T$ (1, 3592) = 164.817, $p < .001$
- $F_{ST}$ (3, 3592) = 16.242, $p < .001$

### P7

**Response times:**
- $F_S$ (3, 3357) = 1.872, $p = .132$
- $F_C$ (1, 3357) = 168.747, $p < .001$
- $F_T$ (1, 3357) = 1273.619, $p < .001$
- $F_{CS}$ (3, 3357) = 0.495, $p = .686$
- $F_{ST}$ (3, 3357) = 0.659, $p = .577$
- $F_{CT}$ (1, 3357) = 23.316, $p < .001$
- $F_{CST}$ (3, 3357) = 6.646, $p < .001$

**Errors:**
- $\chi^2_S$ (3, N=3594) = 1.504, $p = .681$
- $\chi^2_C$ (1, N=3594) = 46.040, $p < .001$
- $\chi^2_T$ (1, N=3594) = 10.288, $p = .001$
- $\chi^2_{CS}$ (3, N=3594) = 3.867, $p = .276$
- $\chi^2_{ST}$ (3, N=3594) = 6.851, $p = .077$
- $\chi^2_{CT}$ (1, N=3594) = 0.383, $p = .536$
- $\chi^2_{CST}$ (3, N=3594) = 13.037, $p = .005$

**Response Accuracy in pID:**
- $\chi^2_S$ (3, N=3600) = 24.471, $p < .001$
- $\chi^2_T$ (1, N=3600) = 15.634, $p < .001$
- $\chi^2_{ST}$ (3, N=3600) = 9.029, $p = .029$

**Ratings on PAS:**
- $F_S$ (3, 3592) = 220.262, $p < .001$
- $F_T$ (1, 3592) = 1827.489, $p < .001$
- $F_{ST}$ (3, 3592) = 50.827, $p < .001$

### P8

**Response times:**
- $F_S$ (3, 3415) = 0.739, $p = .529$
- $F_C$ (1, 3415) = 39.876, $p < .001$
- $F_T$ (1, 3415) = 357.184, $p < .001$
- $F_{CS}$ (3, 3415) = 0.867, $p = .457$
- $F_{ST}$ (3, 3415) = 0.851, $p = .466$
- $F_{CT}$ (1, 3415) = 0.012, $p = .911$
- $F_{CST}$ (3, 3415) = 0.336, $p = .799$

**Errors:**
- $\chi^2_S$ (3, N=3590) = 6.856, $p = .077$
- $\chi^2_C$ (1, N=3590) = 0.000, $p = .995$
- $\chi^2_T$ (1, N=3590) = 73.790, $p = .001$
- $\chi^2_{CS}$ (3, N=3590) = 2.790, $p = .425$
- $\chi^2_{ST}$ (3, N=3590) = 9.432, $p = .024$
- $\chi^2_{CT}$ (1, N=3590) = 0.000, $p = .994$
- $\chi^2_{CST}$ (3, N=3590) = 5.677, $p = .128$

**Response Accuracy in pID:**
- $\chi^2_S$ (3, N=3600) = 1.048, $p = .790$
- $\chi^2_T$ (1, N=3600) = 0.343, $p = .558$
- $\chi^2_{ST}$ (3, N=3600) = 0.425, $p = .935$

**Ratings on PAS:**
- $F_S$ (3, 3592) = 0.800, $p = .494$
- $F_T$ (1, 3592) = 243.045, $p < .001$
- $F_{ST}$ (3, 3592) = 0.906, $p = .437$

**D: Results for individual participants in Experiment 2.**

Table A2 holds the test statistics and *p* values for the individual main effects and interactions reported in graphical form in Fig. 4.

Table A2: Test statistics, degrees of freedom, and *p* values for the main effects and interactions reported in Fig. 4. The large number of denominator degrees of freedom is determined by the number of trials per participant. Note that these analyses aim to generalize to new trials from the same participant rather than to new participants.

|  | **P1** |  |  |  | **P2** |  |  |  | **P3** |  |  |  | **P4** |  |  |
|---|---|---|---|---|---|---|---|---|---|---|---|---|---|---|---|
|  | Response times: |  |  |  | Response times: |  |  |  | Response times: |  |  |  | Response times: |  |  |
| $F_S$ | (3, 3461) | = 3.953 | $p = .008$ | $F_S$ | (3, 3392) | = 2.140 | $p = .093$ | $F_S$ | (3, 3427) | = 1.291 | $p = .276$ | $F_S$ | (3, 3396) | = 26.017 | $p < .001$ |
| $F_C$ | (1, 3461) | = 103.898 | $p < .001$ | $F_C$ | (1, 3392) | = 221.770 | $p < .001$ | $F_C$ | (1, 3427) | = 50.435 | $p < .001$ | $F_C$ | (1, 3396) | = 262.107 | $p < .001$ |
| $F_T$ | (1, 3461) | = 518.197 | $p < .001$ | $F_T$ | (1, 3392) | = 171.702 | $p < .001$ | $F_T$ | (1, 3427) | = 759.111 | $p < .001$ | $F_T$ | (1, 3396) | = 1526.714 | $p < .001$ |
| $F_{CS}$ | (3, 3461) | = 2.796 | $p = .039$ | $F_{CS}$ | (3, 3392) | = 9.755 | $p < .001$ | $F_{CS}$ | (3, 3427) | = 1.833 | $p = .139$ | $F_{CS}$ | (3, 3396) | = 2.304 | $p = .075$ |
| $F_{ST}$ | (3, 3461) | = 10.517 | $p < .001$ | $F_{ST}$ | (3, 3392) | = 0.875 | $p = .453$ | $F_{ST}$ | (3, 3427) | = 0.520 | $p = .668$ | $F_{ST}$ | (3, 3396) | = 16.872 | $p < .001$ |
| $F_{CT}$ | (1, 3461) | = 0.162 | $p = .687$ | $F_{CT}$ | (1, 3392) | = 0.031 | $p = .861$ | $F_{CT}$ | (1, 3427) | = 28.823 | $p < .001$ | $F_{CT}$ | (1, 3396) | = 34.397 | $p < .001$ |
| $F_{CST}$ | (3, 3461) | = 0.982 | $p = .400$ | $F_{CST}$ | (3, 3392) | = 0.789 | $p = .500$ | $F_{CST}$ | (3, 3427) | = 0.861 | $p = .461$ | $F_{CST}$ | (3, 3396) | = 1.420 | $p = .235$ |
|  | Errors: |  |  |  | Errors: |  |  |  | Errors: |  |  |  | Errors: |  |  |
| $\chi^2_S$ | (3, N=3582) | = 4.438 | $p = .218$ | $\chi^2_S$ | (3, N=3583) | = 6.432 | $p = .092$ | $\chi^2_S$ | (3, N=3575) | = 2.768 | $p = .429$ | $\chi^2_S$ | (3, N=3576) | = 2.953 | $p = .399$ |
| $\chi^2_C$ | (1, N=3582) | = 15.388 | $p < .001$ | $\chi^2_C$ | (1, N=3583) | = 29.110 | $p < .001$ | $\chi^2_C$ | (1, N=3575) | = 4.048 | $p = .044$ | $\chi^2_C$ | (1, N=3576) | = 18.499 | $p < .001$ |
| $\chi^2_T$ | (1, N=3582) | = 59.293 | $p < .001$ | $\chi^2_T$ | (1, N=3583) | = 4.075 | $p = .044$ | $\chi^2_T$ | (1, N=3575) | = 31.473 | $p < .001$ | $\chi^2_T$ | (1, N=3576) | = 0.016 | $p = .899$ |
| $\chi^2_{CS}$ | (3, N=3582) | = 1.647 | $p = .649$ | $\chi^2_{CS}$ | (3, N=3583) | = 13.685 | $p = .003$ | $\chi^2_{CS}$ | (3, N=3575) | = 10.961 | $p = .012$ | $\chi^2_{CS}$ | (3, N=3576) | = 2.483 | $p = .478$ |
| $\chi^2_{ST}$ | (3, N=3582) | = 6.602 | $p = .086$ | $\chi^2_{ST}$ | (3, N=3583) | = 4.526 | $p = .210$ | $\chi^2_{ST}$ | (3, N=3575) | = 2.160 | $p = .540$ | $\chi^2_{ST}$ | (3, N=3576) | = 0.156 | $p = .984$ |
| $\chi^2_{CT}$ | (1, N=3582) | = 4.011 | $p = .045$ | $\chi^2_{CT}$ | (1, N=3583) | = 1.661 | $p = .197$ | $\chi^2_{CT}$ | (1, N=3575) | = 1.291 | $p = .256$ | $\chi^2_{CT}$ | (1, N=3576) | = 0.109 | $p = .742$ |
| $\chi^2_{CST}$ | (3, N=3582) | = 1.606 | $p = .658$ | $\chi^2_{CST}$ | (3, N=3583) | = 3.321 | $p = .345$ | $\chi^2_{CST}$ | (3, N=3575) | = 2.478 | $p = .479$ | $\chi^2_{CST}$ | (3, N=3576) | = 1.199 | $p = .753$ |
|  | Response Accuracy in pID: |  |  |  | Response Accuracy in pID: |  |  |  | Response Accuracy in pID: |  |  |  | Response Accuracy in pID: |  |  |
| $\chi^2_S$ | (3, N=3584) | = 12.962 | $p = .005$ | $\chi^2_S$ | (3, N=3584) | = 4.315 | $p = .229$ | $\chi^2_S$ | (3, N=3584) | = 97.064 | $p < .001$ | $\chi^2_S$ | (3, N=3584) | = 10.635 | $p = .014$ |
| $\chi^2_T$ | (1, N=3584) | = 206.661 | $p < .001$ | $\chi^2_T$ | (1, N=3584) | = 155.259 | $p < .001$ | $\chi^2_T$ | (1, N=3584) | = 611.273 | $p < .001$ | $\chi^2_T$ | (1, N=3584) | = 67.220 | $p < .001$ |
| $\chi^2_{ST}$ | (3, N=3584) | = 10.331 | $p = .016$ | $\chi^2_{ST}$ | (3, N=3584) | = 8.496 | $p = .037$ | $\chi^2_{ST}$ | (3, N=3584) | = 94.087 | $p < .001$ | $\chi^2_{ST}$ | (3, N=3584) | = 10.463 | $p = .015$ |
|  | **P5** |  |  |  | **P6** |  |  |  | **P7** |  |  |  | **P8** |  |  |
|  | Response times: |  |  |  | Response times: |  |  |  | Response times: |  |  |  | Response times: |  |  |
| $F_S$ | (3, 3465) | = 0.241 | $p = .868$ | $F_S$ | (3, 3344) | = 1.678 | $p = .170$ | $F_S$ | (3, 3410) | = 1.548 | $p = .200$ | $F_S$ | (3, 3366) | = 0.850 | $p = .467$ |
| $F_C$ | (1, 3465) | = 11.110 | $p = .001$ | $F_C$ | (1, 3344) | = 191.854 | $p < .001$ | $F_C$ | (1, 3410) | = 5.134 | $p = .024$ | $F_C$ | (1, 3366) | = 422.049 | $p < .001$ |
| $F_T$ | (1, 3465) | = 788.574 | $p < .001$ | $F_T$ | (1, 3344) | = 142.569 | $p < .001$ | $F_T$ | (1, 3410) | = 662.776 | $p < .001$ | $F_T$ | (1, 3366) | = 703.736 | $p < .001$ |
| $F_{CS}$ | (3, 3465) | = 2.673 | $p = .046$ | $F_{CS}$ | (3, 3344) | = 9.227 | $p < .001$ | $F_{CS}$ | (3, 3410) | = 3.548 | $p = .014$ | $F_{CS}$ | (3, 3366) | = 7.389 | $p < .001$ |
| $F_{ST}$ | (3, 3465) | = 0.788 | $p = .501$ | $F_{ST}$ | (3, 3344) | = 1.804 | $p = .144$ | $F_{ST}$ | (3, 3410) | = 4.974 | $p = .002$ | $F_{ST}$ | (3, 3366) | = 0.953 | $p = .414$ |
| $F_{CT}$ | (1, 3465) | = 0.220 | $p = .639$ | $F_{CT}$ | (1, 3344) | = 53.833 | $p < .001$ | $F_{CT}$ | (1, 3410) | = 3.288 | $p = .070$ | $F_{CT}$ | (1, 3366) | = 158.951 | $p < .001$ |
| $F_{CST}$ | (3, 3465) | = 5.558 | $p = .001$ | $F_{CST}$ | (3, 3344) | = 2.694 | $p = .045$ | $F_{CST}$ | (3, 3410) | = 0.421 | $p = .738$ | $F_{CST}$ | (3, 3366) | = 3.441 | $p = .016$ |
|  | Errors: |  |  |  | Errors: |  |  |  | Errors: |  |  |  | Errors: |  |  |
| $\chi^2_S$ | (3, N=3578) | = 11.271 | $p = .010$ | $\chi^2_S$ | (3, N=3584) | = 8.433 | $p = .038$ | $\chi^2_S$ | (3, N=3580) | = 0.971 | $p = .808$ | $\chi^2_S$ | (3, N=3581) | = 1.837 | $p = .607$ |
| $\chi^2_C$ | (1, N=3578) | = 0.807 | $p = .369$ | $\chi^2_C$ | (1, N=3584) | = 20.560 | $p < .001$ | $\chi^2_C$ | (1, N=3580) | = 0.006 | $p = .939$ | $\chi^2_C$ | (1, N=3581) | = 2.770 | $p = .096$ |
| $\chi^2_T$ | (1, N=3578) | = 15.699 | $p < .001$ | $\chi^2_T$ | (1, N=3584) | = 0.155 | $p = .694$ | $\chi^2_T$ | (1, N=3580) | = 23.944 | $p < .001$ | $\chi^2_T$ | (1, N=3581) | = 11.340 | $p = .001$ |
| $\chi^2_{CS}$ | (3, N=3578) | = 5.107 | $p = .164$ | $\chi^2_{CS}$ | (3, N=3584) | = 18.369 | $p < .001$ | $\chi^2_{CS}$ | (3, N=3580) | = 6.478 | $p = .091$ | $\chi^2_{CS}$ | (3, N=3581) | = 2.436 | $p = .487$ |
| $\chi^2_{ST}$ | (3, N=3578) | = 5.360 | $p = .147$ | $\chi^2_{ST}$ | (3, N=3584) | = 2.778 | $p = .427$ | $\chi^2_{ST}$ | (3, N=3580) | = 2.780 | $p = .427$ | $\chi^2_{ST}$ | (3, N=3581) | = 1.139 | $p = .768$ |
| $\chi^2_{CT}$ | (1, N=3578) | = 1.535 | $p = .215$ | $\chi^2_{CT}$ | (1, N=3584) | = 14.989 | $p < .001$ | $\chi^2_{CT}$ | (1, N=3580) | = 3.885 | $p = .049$ | $\chi^2_{CT}$ | (1, N=3581) | = 0.735 | $p = .391$ |
| $\chi^2_{CST}$ | (3, N=3578) | = 5.509 | $p = .138$ | $\chi^2_{CST}$ | (3, N=3584) | = 2.160 | $p = .540$ | $\chi^2_{CST}$ | (3, N=3580) | = 2.366 | $p = .500$ | $\chi^2_{CST}$ | (3, N=3581) | = 10.481 | $p = .015$ |
|  | Response Accuracy in pID: |  |  |  | Response Accuracy in pID: |  |  |  | Response Accuracy in pID: |  |  |  | Response Accuracy in pID: |  |  |
| $\chi^2_S$ | (3, N=3584) | = 29.170 | $p < .001$ | $\chi^2_S$ | (3, N=3584) | = 62.129 | $p < .001$ | $\chi^2_S$ | (3, N=3584) | = 310.401 | $p < .001$ | $\chi^2_S$ | (3, N=3584) | = 95.535 | $p < .001$ |
| $\chi^2_T$ | (1, N=3584) | = 24.027 | $p < .001$ | $\chi^2_T$ | (1, N=3584) | = 6.017 | $p = .014$ | $\chi^2_T$ | (1, N=3584) | = 21.568 | $p < .001$ | $\chi^2_T$ | (1, N=3584) | = 0.34 | $p = .853$ |
| $\chi^2_{ST}$ | (3, N=3584) | = 1.696 | $p = .638$ | $\chi^2_{ST}$ | (3, N=3584) | = 6.190 | $p = .103$ | $\chi^2_{ST}$ | (3, N=3584) | = 9.352 | $p = .025$ | $\chi^2_{ST}$ | (3, N=3584) | = 1.021 | $p = .796$ |

**Footnotes**

[1] Because the same stimulus serves as mask and target in the present experiments, we refer to it simply as *mask* in this paper. In general, response priming can be employed with or without masking, the mask can be the target or a separate stimulus, flankers can be used for primes or for targets, and two-alternative forced choice tasks can be employed instead of two-interval yes-no tasks. There is a continuum of variation between typical response priming and typical Eriksen flanker tasks, which are probably variants of the same effect (F. Schmidt et al., 2011).

[2] In experiments on response priming, we would usually use an upper cutoff value of 999 ms, which would lie in the far right tail of the RT distribution. Here, we adjusted this criterion to 1199 ms based on visual inspection of the RT histogram to cut a similar proportion as usual.

[3] These findings support the idea of a preselection into type-A and type-B observers on the basis of a pilot session, a method that has been used successfully by Albrecht and colleagues (e.g., Albrecht & Mattler, 2012; Albrecht, Klapötke, & Mattler, 2010).

[4] One way the observer could harvest additional information about the prime in the dual task would be to notice signs of the response conflict in the response to the mask and use this to guess the identity of the prime. Errors in mask ID would be especially diagnostic of inconsistent primes. However, prime ID accuracy was lower, not higher, on error trials than on correct trials, and so we have to exclude this potential strategy.

**Open practices statement**

All data, materials and analyses are available from the authors upon request. All independent and dependent variables are identified. Preregistration was not employed.

**Funding**

This research did not receive any specific grant from funding agencies in the public, commercial, or not-for-profit sectors.

**Author note**

We thank our participants, our colleagues Maximilian Wolkersdorfer and Sven Panis, and our research assistants. Last but not least, we especially thank our bachelor students Finja Coerdt, Julia Hapke, Lena Schröer, Philipp Neuer, and Simone Eberts for data collection.

**Preprint note**

This preprint is a preliminary version of a manuscript still under review. It is still subject to changes, and reviewer suggestions are still going to be incorporated.